\documentclass[12pt,preprint]{aastex}

\makeatletter

\newcommand{\Rmnum}[1]{\expandafter\@slowromancap\romannumeral #1@}
\makeatother

\slugcomment{Not to appear in Nonlearned J., 45.}

\shorttitle{Circumbinary Terrestrial Planets Formation}
\shortauthors{Gong, Zhou, \& Xie}

\begin{document}


\title{Terrestrial Planets Formation around Circumbinary Habitable Zone: Inward  Migration in the Planetesimal Swarm
}


\author{Yan-Xiang Gong\altaffilmark{1,2}, Ji-Lin Zhou\altaffilmark{1}, Ji-Wei Xie\altaffilmark{1}}
\affil{$^{1}$ Department of Astronomy \& Key Laboratory of Modern Astronomy and Astrophysics in Ministry of Education, Nanjing University,
    Nanjing, 210093, China}
\affil{$^{2}$ College of Physics and Electronic Engineering, Taishan University, Taian, 271021, China}
\email{yxgong@nju.edu.cn, zhoujl@nju.edu.cn}



\begin{abstract}
According to the core accretion theory, circumbinary embryos can form only beyond a critical semimajor axis (CSMA).
However, due to the relatively high density of solid
materials in the inner disk, significant amount of small planetesimals must exist in the inner zone when
embryos were forming outside this CSMA. So embryos migration induced by the planetesimal swarm
is possible after the gas disk depletion. Through numerical simulations, we found (i) the scattering-driven inward migration of embryos is robust,
planets can form in the habitable zone if we adopt a mass distribution of MMSN-like disk;
(ii) the total mass of the planetesimals in the inner region and continuous embryo-embryo scattering are two key factors that cause significant embryo migrations;
(iii) the scattering-driven migration of embryos is a natural water-deliver mechanism. We propose that planet detections should focus on the close
binary with its habitable zone near CSMA.

\end{abstract}


\section{Introduction}

Circumbinary planets - planets encircle both members of a binary are interesting  celestial bodies discovered by exoplanet-hunting missions.
So far, six exoplanets circling main-sequence
binary stars were found by Kepler mission, they are Kepler-16 b \citep{doyle11}, Kepler-34 b, Kepler-35 b \citep{welsh12}, Kepler-38 b \citep{orosz12a}, Kepler-47 b and Kepler-47 c \citep{orosz12b}. Though exoplanet-hunting plans selectively avoid close binaries, circumbinary planets may be common. Researches manifest that more than $1\%$ of close binary stars have giant
planets in nearly coplanar orbits, which yielding a Galactic population of at least several million \citep{welsh12}.
All the six circumbinary planets discovered hitherto by Kepler are Saturn-like or Neptune-like planets, they are not suitable for the existence of extraterrestrial life.
We know the ultimate aim of Kepler is to find Earth-like exoplanets, and to search for extrasolar life. It is natural to
ask whether terrestrial planets can form in the circumbinary case? This {\em Letter} intends to explore some aspects of this topic.

The semimajor axis (SMA), eccentricity and mass ratio of
a binary are designated by $a_{B}$, $e_{B}$ and $\mu=M_{B}/(M_{A}+M_{B})$ ($M_A$ is the bigger one) in the paper, respectively. According to \citet{holman99},
there is a critical SMA ($a_{c1}$) within which test particles are unstable (beside some specific initial configurations).
Based on the core accretion model, there is another critical SMA ($a_{c2}$, usually outside of $a_{c1}$) within which planetesimal accretion is impossible
due to the strong perturbation of the binary.
In this Letter, we propose a mechanism to form terrestrial planets in the habitable zone (HZ) through inward migration of embryos outside $a_{c2}$.
It is organized as follows: a summary of former works and our ideas are presented in Section 2;  Session 3 and 4 show our model and simulation results. The conclusions are summarized in Section 5.

\section{Synopsis of former works and our ideas}

{\em From planetsimals to embryos}.
In the circumbinary case, the perturbations of the companion star can stimulate the eccentricities of the planetesimals, which lead to high
 impact velocities between planetesimals.
By ignoring the effect of gas,
\citet{moriwaki04} concluded that the planetesimal
accretion cannot occur within 13 AU ($\mu=0.2$, $a_{B}=1$, $e_{B}=0.1$).
The coupled effects of companion perturbations  and gas friction lead to strong size-dependent orbital phasing,  which reduces the planetesimal impact velocities and makes  accretion  more optimistic \citep{scholl07}.
However, when the disk evolution is included,  pericenters of planetesimals are less collimated than
that in the stationary and axisymmetric disk model, which again inhibits the planetesimal accretion \citep{marzari08}.

As a result, \citet{meschiari12a} concluded planetesimal accretion might be effectively inhibited within 4 AU in Kepler-16(AB).
\citet{meschiari12b} considered the turbulence driven by the magnetorotation instability.
He found that planetesimal accretion can be inhibited even in 4-10 AU region in Kepler-16(AB).
\citet{paardekooper12}
confirmed that {\it in situ} growth starting from planetesimals smaller than $\sim 10$ km is difficult for Kepler 16b, Kepler 34b, and Kepler 35b.

{\em From the embryos to the terrestrial planets}. \citet{quintana06}  found the apastron distance $Q_B = a_{B}(1 + e_{B})$ is a critical parameter
for the formation of circumbinary terrestrial planets. When $Q_{B} < 0.2$ AU (for $0.05\;{\rm{AU}} \le {a_B} \le 0.4\;{\rm{AU}}$),
the formed planetary systems are very similar to those around single stars, whereas those with larger
$Q_{B}$ tend to harbour fewer planets, especially interior to 1 AU. However, they didn't consider the origin of the initial embryos.

Since circumbinary embryos can only form in large distances (beyond the ice line in Kepler-16 system \citep{meschiari12a}),
embryos' inward migration is the only path to form terrestrial planets in the inner zone.
There are two kinds of migrations: type I migration driven by gas disk \citep{Gold79, Ward97} and
the migration of embryos driven by the planetesimal disk \citep{ida00}.
\citet{pierens07, pierens08} have shown that planets with terrestrial to Saturn mass always migrate
inwards till they reach the inner border of the disk truncated by the tidal force of the binary.
However, from the viewpoint of planet formation, there are some inevitable problems for
the formation of the circumbinary embryos or protoplanets.
\begin{itemize}
\item Long accretion timescales of distant embryos.
Outside $a_{c2}$,  planetesimal accretion  will be slower than that in single-star systems due to  the perturbations of the companion star,
with the so-called type II runaway growth \citep{kortenkamp01}. On the other hand, $a_{c2}$ is a function of time, over longer
timescales, the impact velocities in the outer regions will also increase, which makes accretion  more slowly \citep{thebault06}.
\item Rapid dissipation of the circumbinary gas disk. Simulations of \citet{artymowicz94} and \citet{fragner10} manifest that the
disk will be truncated, distorted and teared quickly, so the the lifetime of the circumbinary gas disk will be shorter.
Some evidences for the reduced lifetimes ($\sim$ 0.1-1 Myr) of the primordial circumstellar disk in binary system had been confirmed by observations \citep{cieza09}.
\item Uncertainty of the type I migration. Recent studies show that the orientation and migration rate of type I migration is very uncertain \citep{kley12, ormel12}. Inward migration
can not be guaranteed in all kinds of disk environments. A notable example is the solar system where planets might not have obvious gas-driven migration.\footnote{A new point can be found in \citet{walsh12}.} On the other hand, type I migration is unimportant for the terrestrial planets \citep{bromley11}. The reason is that building blocks of solid planets are packed too closely to migrate.
\end{itemize}
Long accretion timescales of embryos and the reduced lifetime of gas disks indicate that significant gas-disk migration of circumbinary planet {\it may be} less significant as in single star systems. However, due to the relatively high density of solid materials in the inner disk, significant amount of small planetesimals must exist in the inner region when small embryos was forming outside of $a_{c2}$. Thus planetesimal-driven migration of embryos maybe work. In this paper, we consider the migration and accretion of embryos in a planetesimal disk---a `gas free' case that the influence of the residual gas disk can be ignored. Our work can be thought as a complementary work of \citet{pierens07, pierens08}---a case that gas disk has dissipated when the embryos formed outside of $a_{c2}$. The aim is to check whether the terrestrial planets can form in the inner region such as habitable zone (HZ) under such `gas free' circumstance.  \citet{payne09} have discussed the migration of S-type embryos in a planetesimal disk, we apply a similar scattering-driven migration mechanism in the circumbinary case.

\section{Numerical Model and Initial Conditions}

The parameter space of the binary systems is relatively large. Here, we select the Kepler-16(AB) as the fiducial binary configuration.
The fitting formula found in \citet{holman99} is used to calculate the unstable boundary $a_{c1}$ (error bar is ignored),
\begin{equation}\label{ac1}
\begin{array}{*{20}{l}}
{{a_{c1}} = \left[ {1.60 + 5.1{e_B} - 2.22e_B^2 + 4.12\mu  - } \right.}\\
{\;\;\;\;\;\;\;\left. {4.27{e_B}\mu  - 5.09{\mu ^2} + 4.61e_B^2{\mu ^2}} \right]{a_B}.}
\end{array}
\end{equation}
Using the parameter of Kepler-16(AB) (see Table 1), we get $a_{c1}\approx0.66$ AU. The $a_{c2} = 4$ AU derived by \citet{meschiari12a} is used to define the boundary inside which the
embryos formation is effectively inhibited. We call $a_{c1}$ to $a_{c2}$ the inner region and $>a_{c2}$ the outer region.
According to the `minimum-mass solar nebula' (MMSN) disk \citep{hayashi81}, the mass of solids within the inner region is about 6-8 $M_{\oplus}$ (scaled by $(M_{A}+M_{B})/M_\odot$).
We suppose the total mass of planetesimals in the inner region is 6 $M_{\oplus}$ in our model. Planetesimals are randomly distributed according to the $a^{-\alpha}$ profile.
300 planetesimals are put in the inner region, each with a mass of 0.02 $M_{\oplus}$.
Outside  $a_{c2}$,  we assume the runaway growth has been taken place, resulting a chain of isolated embryos. Because the final mass
profile of embryos is unclear in the circumbinary case, we take the $\alpha$-disk model given by \citet{kokubo02}, and explore
six kinds of possible value of $\alpha$ ($\alpha$ = 0.0, 0.5, 1.0, 1.5, 2.0, 2.5).
The distance between embryos is 7 mutual Hill radius. We put 100 embryos in the outer region following \citet{kokubo02}, namely,
\begin{equation}
\begin{array}{l}
{M_{{\rm{iso}}}} \simeq 0.16{\left( {\frac{{\tilde b}}{{10}}} \right)^{{3 \mathord{\left/
 {\vphantom {3 2}} \right.
 \kern-\nulldelimiterspace} 2}}}{\left( {\frac{{{f_{{\rm{ice}}}}{\Sigma _1}}}{{10}}} \right)^{{3 \mathord{\left/
 {\vphantom {3 2}} \right.
 \kern-\nulldelimiterspace} 2}}}\\
\,\,\,\,\,\,\,\,\,\,\,\,\,\, \times {\left( {\frac{a}{{1\,{\rm{AU}}}}} \right)^{\left( {{3 \mathord{\left/
 {\vphantom {3 2}} \right.
 \kern-\nulldelimiterspace} 2}} \right)\left( {2 - \alpha } \right)}}{\left( {\frac{{{M_ * }}}{{{M_ \odot }}}} \right)^{{{ - 1} \mathord{\left/
 {\vphantom {{ - 1} 2}} \right.
 \kern-\nulldelimiterspace} 2}}}\,\,{M_ \oplus },
\end{array}
\end{equation}
\begin{equation}
\begin{array}{l}
{r_{\rm{H}}} \simeq 0.69 \times {10^{ - 2}}{\left( {\frac{{\tilde b}}{{10}}} \right)^{{3 \mathord{\left/
 {\vphantom {3 2}} \right.
 \kern-\nulldelimiterspace} 2}}}{\left( {\frac{{{f_{{\rm{ice}}}}{\Sigma _1}}}{{10}}} \right)^{{1 \mathord{\left/
 {\vphantom {1 2}} \right.
 \kern-\nulldelimiterspace} 2}}}\\
\,\,\,\,\,\,\,\,\,\,\,\,\,\, \times {\left( {\frac{a}{{1\,{\rm{AU}}}}} \right)^{\left( {{1 \mathord{\left/
 {\vphantom {1 2}} \right.
 \kern-\nulldelimiterspace} 2}} \right)\left( {4 - \alpha } \right)}}{\left( {\frac{{{M_ * }}}{{{M_ \odot }}}} \right)^{{{ - 1} \mathord{\left/
 {\vphantom {{ - 1} 2}} \right.
 \kern-\nulldelimiterspace} 2}}}\,\,{\rm{AU}},
\end{array}
\end{equation}
where $\tilde b$ is the orbital separation of embryos (scaled by the mutual Hill radius), $\Sigma_{1}$ is the reference surface density at 1 AU, $f_{ice}$ is the ice factor and $M_*$ is the mass of the central stars.
Because all the embryos lie outside the ice line of the Kepler-16(AB) ($\sim$2.3 AU) \citep{meschiari12a},
$f_{ice}\Sigma_{1}\equiv \Sigma_{ice}$ can be treated as a single parameter. As we only focus on the formation of Earth-like planets,
the mass of embryos is limited between 0.01 to 0.1 $M_{\oplus}$ (Moon-like to Mars-like mass). It can be obtained by carefully adjusting
the parameter $\Sigma_{ice}$, resulting the total mass of embryos $\sim 6 M_{\oplus}$ in the outer region (4.0 AU to 6.6 AU). We found the total mass of embryos in this region can be compatible with the MMSN disk ($\sim 7 M_{\oplus}$ in this region).

We integrate the system to $10^{6}$ yr. The Bulirsch-Stoer integrator in MERCURY package \citep{chambers99} is modified to model the close binary
configuration. The accuracy parameter of $10^{-12}$ is used, all integrations conserved total energy and angular momentum within $10^{-6}$.

\section{Simulation Results and Analysis}

\subsection{The Fiducial Case} \label{bozomath}


We take an MMSN disk ($\alpha$ = 1.5) as the fiducial case. Planetesimals are simulated as test particles (with mass), whereas embryos are modeled as fully interacting bodies.
Because scattering is a chaotic process, we perform 20 simulations for such a disk. The angular orbital elements of planetesimals and embryos
are randomly selected. A representative outcome of the fiducial case is given in Fig. 1. After 1 Myr, the original 3 embryos outside $a_{c2}$ have migrated
into the inner region. The innermost one has a SMA of 0.95 AU, the mass is $M_{im}=0.89\,M_{\oplus}$, and the eccentricity is 0.037.
To compare with different runs, we define a migration rate as
\begin{equation}
\eta \left( t \right) = \frac{{{a_{im,\,f}} - {a_{im,\,\,i}}}}{{{a_{im,\,\,i}}}},
\end{equation}
where $a_{im,\,f}$ is the final SMA of the innermost embryo and $a_{im,\,i}$ is the initial SMA of the same embryo.
We find $\eta = - 0.75_{ - 0.04}^{ + 0.05}$ for the 20 runs, where the errors denote the upper and lower quartiles.

In Fig. 2, we give migration rate as a function of time, averaged over 20 simulations. We find that, after
$2\times10^{5}$ yr ($\eta = - 0.69_{ - 0.04}^{ + 0.04}$), no significant migration takes place, which indicates that
the system has reached an equilibrium state, thus the statistical results are reliable.
In the fiducially model, we find 2 out of 20 runs (so, $10\%$ chance)
result in a planet in the HZ of Kepler-16(AB) (Table 1).

\subsection{Parameter Exploration} \label{bozomath}
{\em Total mass of embryos ($M_{tot}$)}.
We vary the number of embryos outside and keeping other parameters unchanged. $N_{e}$=200, 150, 100, 50, 25 are explored. Moreover, an extreme case $N_{e}$=1 is also studied. We conduct 20 runs for each $N_e$, results are shown in Table 2.
We find $N_{e}$ has no meaningful effect on the migration rate and the mass of innermost embryo as long as $N \geq 100$. But when $N \leq 25$, migration rate decreases significantly. At the extreme case $N_e=1$, embryo has negligible migration.

{\em Location of CSMA ($a_{c2}$)}. The initial size and distribution of planetesimals, the value of the critical accretion velocity, the growth of planetesimals and other factors have great uncertainty in determining  $a_{c2}$.  Different models give different $a_{c2}$. E.g., \citet{marzari12} shows $a_{c2}\sim 3$ AU,  while  \cite{paardekooper12} find $a_{c2}\geq4.4$ AU for Kepler 16(AB). Here, $a_{c2}$ is varied to check its effect on migration rate. We consider $a_{c2}$=[3.0, 3.5, 4.0, 4.5, 5.0] AU.
$\Sigma_{ice}$ is adjusted to keep the total mass of embryos and planetesimals unchanged (equal to the fiducial case). The results are shown in Fig. 3a, we find no significant changes in the migration rate and the masses of the innermost planets that formed.

{\em Density profile ($\alpha$)}. The material distribution of a circumbinary disk is unclear. We consider other five kinds of density profiles: $\alpha=$0.0, 0.5, 1.0, 2.0, 2.5. The results are shown in Fig. 3b.
Again, no significant change in the migration rate and mass, which manifests that the inward migration of embryos is generally robust.

{\em Total mass of planetesimals}. Here we consider two extreme cases only: 1, no planetesimals in the inner region; 2, the total mass of planetesimals is double that of fiducial case (ice line is in the inner region). Other parameters are unchanged. We find that in the absence of planetesimals, migration of the embryos are severely suppressed, generally with $a_{im,f}>$3 AU. For example, if we remove the planetesimals in Fig. 1, the SMA of innermost planet is $\sim$3.5 AU (see right panel of Fig. 1). It shows that significant (inward) migration is caused not only by the scattering between embryos but also by planetesimals disk. If the mass of planetesimals is doubled, we have $\eta=-0.76_{-0.04}^{+0.07}$, $a_{im}=0.90$ AU. Increasing the total mass of the planetesimals causes stronger inward migration (compared to the fiducial case). The probability of forming a HZ planet is also increased ($20\%$).

{\em The number of planetesimals ($N_t$)}. In order to improve the computational efficiency, we model the planetesimals as test particles with associated mass.
To validate that this is a reasonable approximation, we perform additional two suits of runs with  $N_t$=500, 1000, the total mass of the planetesimas is unchanged. The results are $\eta=-0.74_{-0.05}^{+0.03}$; $\eta=-0.73_{-0.04}^{+0.08}$, respectively. Obviously, using a larger number of planetesimals gives us the similar results.

\subsection{Other Systems}

We explore the embryos migration in the other two binary configurations: Kepler-34(AB) and Kepler-35(AB). We use a sample formula to estimate the boundary of circumbinary HZ,
\begin{equation}
\left[ {\rm{H{Z_{in}}},\;\;\rm{H{Z_{out}}}} \right] = \sqrt {\frac{{{L_A} + {L_B}}}{{{L_ \odot }}}}  \cdot \left[ {0.95,\;\;2.0} \right]\;\;{\rm{AU}}.
\end{equation}
$\left[ {0.95,\;\;2.0} \right]$ AU is the extended HZ of the sun \citep{kasting93, mischna00, quarles12}. $L_{A}+L_{B}$ and $L_{\odot}$ are luminosities of the binary and the sun. By including planetesimal formation and dust accretion, \citet{paardekooper12} find $a_{c2}$= 2.76 AU for Kpeler-34(AB) and $a_{c2}$=2.7 AU for Kepler-35(AB). The location of the ice line for an irradiated disk can be estimated from the scaling $2.7{\left[ {\left( {{M_A} + {M_B}} \right)/{M_ \odot }} \right]^2}$ AU \citep{meschiari12a}. We get 10 AU for Kepler-34(AB), and 7.8 AU for Kepler-35(AB), so we can ignore the material enhancement caused by the ice line in putting embryos due to $a_{ice} > a_{c2}$. Other initial conditions have an analogy to the fiducial case of Kepler-16(AB). Results are summarized in Table 1. Again, we find that formation of planets in HZ is possible.

\subsection{The effects of a pre-existing giant planet}
Three giant planets discovered in Kepler-16(AB), -34(AB) and -35(AB) lie just outside the stability boundaries of the binaries
\citep{meschiari12a}. In order to check how a pre-existing planet might affect the previous migration scenario, we performed several additional simulations by initially placing a planet in orbits
similar to Kepler-16 b, -34 b, -35 b. Other initial conditions are similar to the fiducial case. We find $\eta = - 0.70_{ - 0.08}^{ + 0.07}$ and $a_{min}=1.15$ AU in Kepler-16(AB).
No innermost planets can always stay in HZ (with $a_{p}(1+e_{p})> \rm{H{Z_{out}}}$). But in Kepler-34(AB),
we get $\eta = - 0.44_{ - 0.03}^{ + 0.06}$ and $a_{min}=1.57$ AU, 75$\%$ innermost planet are located in HZ.
Kepler-35(AB) have about 90$\%$ innermost planet in HZ. Giant planet and binary compose a three-body system,
it has a new stable boundary $a_{c3}$ (larger than $a_{c1}$), the planetesimal disk that we used in the fiducial case is
truncated by this $a_{c3}$. As a result, the embryo migration is limited by $a_{c3}$. Actually,
if no giant planets are included in Kepler-34(AB) and -35(AB),
some innermost planets migrate over HZ, so the fraction of the innermost planet in the HZ (Table 1) is smaller than the case with a giant planet.

\section{Summary}
Inward migration of the circumbinary embryos driven by planetesimal swarm is discussed in this Letter, the main results and their implications are summarized as follows.
(i) Scattering-driven {\em inward} migration of embryos is robust in the circumbinary case. In the three close binary configurations we explored,
planets can form in the HZ if we adopt mass distributions of MMSN-like disk.
(ii) The total mass of the planetesimals in the inner region is the main factor that affects the migration (rate).
Embryo-embryo scattering is another necessary factor that causes significant migration. Migration of a single embryo
is limited because its increasing mass will reduce the migration rate, ultimately, it clears the adjacent planetesimals, migration stops.
While in a multiple-embryo system, embryo-embryo scattering can continually push the smaller embryo inward.
(iii) We propose that planet detection should focus on the close binary with its HZ near $a_{c2}$.
(iv) Scattering-driven migration of embryos is a nature water-deliver mechanism.
In the close binary like Kepler-16(AB), embryos can only form outside of the snow line ($a_{c2}>a_{ice}$), so the planet found in the
HZ should have water contents.

\acknowledgments

We thank the anonymous referee for his constructive comments
and suggestions. This work is supported by National Basic Research Program of China (973 Program 2013CB834900), NSFC (10925313, 10833001), National Basic Research Program of
China, the Fundamental Research Fund for Central Universities (1112020102), and a Fund from the Chinese Ministry of Education. Gong Yan-Xiang
also acknowledge the support form Shandong Provincial Natural Science Foundation, China (No. ZR2010AQ023, ZR2010AM024).

\clearpage


\begin{table}
\begin{center}
\caption{Binaries parameters and a summary of simulation results.\label{tbl-1}}
\begin{tabular}{llll}
\tableline\tableline
Binary                        & Kepler-16(AB)$^{\rm{a}}$                    & Kepler-34(AB)$^{\rm{b}}$       &  Kepler-35(AB)$^{\rm{b}}$ \\
\tableline
$M_A$ (${M_\odot}$)           & 0.6897                   & 1.0479      & 0.8877    \\
$M_B$ (${M_\odot}$)           & 0.2026                   & 1.0208       & 0.8094\\
$a_B$ (AU)                     &0.22431                    & 0.22882    & 0.17617 \\
$e_B$                             &0.15944                    &0.52087  &0.1421\\
$L_{A}+L_{B}$ ($L_{\odot}$)       &--                     & 2.77            & 1.35       \\
$a_{c1}$ (AU)                 &0.66                  &0.84   & 0.50 \\
$a_{c2}$ (AU)                  &4.00$^{\rm{c}}$                    &2.76$^{\rm{d}}$    &2.70$^{\rm{d}}$ \\
$a_{ice}$ (AU)                  &2.3                    &10.0     &7.8   \\
HZ (AU)                       &  $0.36 \sim 1.02$$^{\rm{e}}$      & $1.58 \sim 3.33$   & $1.10 \sim 2.32$ \\
$\eta$                            &$-0.75_{-0.04}^{+0.05}$    &$-0.49_{-0.11}^{+0.07}$  &$-0.46_{-0.09}^{+0.11}$  \\
$M_{im}$ ($M_\oplus$)       &$0.98_{-0.20}^{+1.12}$    &$1.19_{-0.32}^{+0.37}$                    & $1.18_{-0.28}^{+0.63}$ \\
$a_{min}$ (AU)                 &0.95                   &1.33             & 1.09 \\
EIHZ (Fraction)        &$10\%$                    &$35\%$               &$20\%$ \\
\tableline
\end{tabular}
\tablenotetext{a}{\citet{doyle11}; $^{\rm{b}}$\citet{welsh12}; $^{\rm{c}}$\citet{meschiari12a}; $^{\rm{d}}$\citet{paardekooper12}; $^{\rm{e}}$\citet{quarles12}}
\tablecomments{$a_{min}$ is the minimum final SMA of the embryos in 20 runs. EIHZ means the fraction of the innermost planet in the habitable zone.}
\end{center}
\end{table}

\begin{table}
\begin{center}
\caption{Numerical results of changing the number of embryos.\label{tbl-2}}
\begin{tabular}{lcccccc}
\tableline
\tableline
$N_e$                          &200      &150    &100   &50    &25    &1    \\
\tableline
${M_{tot}}$ (${M_ \oplus }$) &14.72     &9.81   &5.87  &2.65   &1.27   &0.05    \\
$\eta$                         &$-0.76_{-0.10}^{+0.15}$   & $-0.75_{-0.07}^{+0.10}$  &$-0.75_{-0.04}^{+0.05}$  &$-0.62_{-0.03}^{+0.06}$   &$-0.50_{-0.04}^{+0.07}$   &$-0.06_{-0.02}^{+0.06}$   \\
$M_{im}$ (${M_ \oplus }$)   &$1.06_{-0.20}^{+1.14}$   & $1.65_{-0.65}^{+0.99}$  &$0.98_{-0.20}^{+1.15}$  &$0.91_{-0.07}^{+0.35}$   &$1.14_{-0.11}^{+0.11}$   &$0.45_{-0.10}^{+0.02}$   \\
$a_{min}$ (AU)             &0.93     &0.94   &0.95  &1.39   &1.57   &3.36    \\
\tableline
\end{tabular}
\end{center}
\end{table}

\begin{figure}
\epsscale{1.0}
\plotone{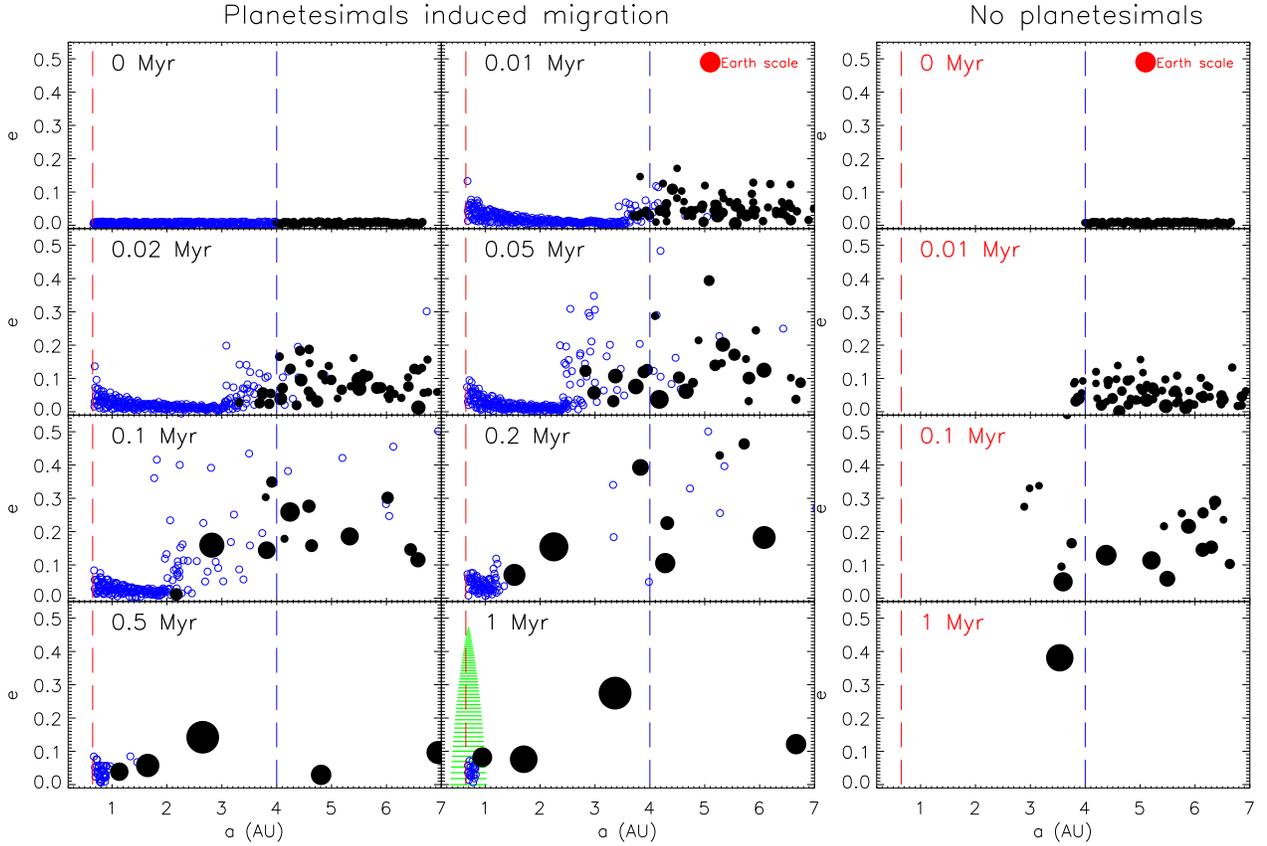}
\caption{The time evolution of the circumbinary planetsimals and embryos. The binary is the Kepler-16(AB) configuration (see Table 1). The embryos are represented by filled circles whose size are proportional to the physics sizes of the bodies (in scale relative to the Earth). The planetesimals are represented by open circles but not in scale relative to the Earth. The dashed red line denotes $a_{c1}$ and the dashed blue line denotes $a_{c2}$. The green region is the habitable zone (0.36 AU $\le {a_p}\left( {1 - {e_p}} \right)$ and ${a_p}\left( {1 + {e_p}} \right) \le 1.02$ AU). Right panel: time evolution of the circumbinary embryos if the initial planetesimal swarm is removed.
\label{fig1}}\end{figure}
\clearpage

\begin{figure}
\epsscale{0.8}
\plotone{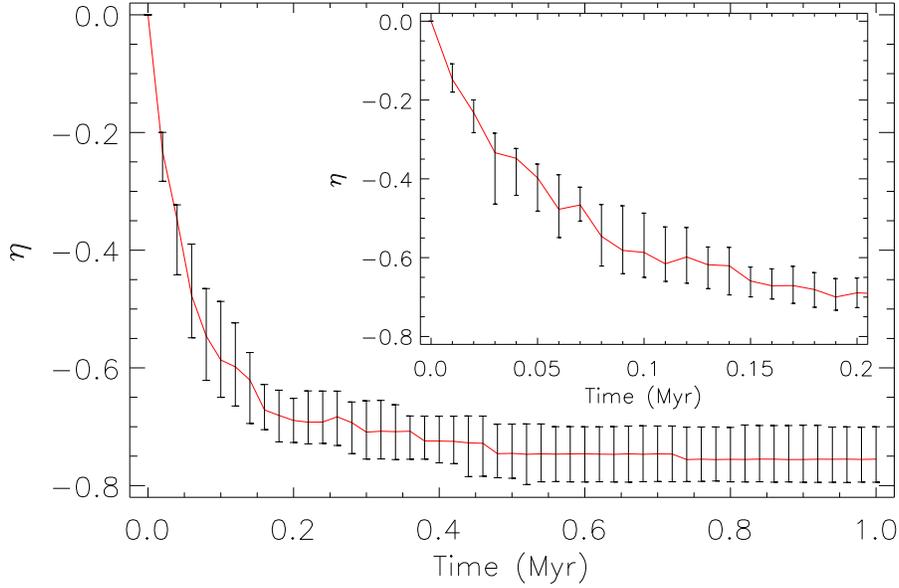}
\caption{The value of $\eta=(a_{im,\,f} - a_{im,\,\,i})/a_{im,\,\,i}$ as a function of time, averaged over 20 runs. The median value is plotted, as well as the lower and upper quartile bounds.
The inline image is zoom of the first 0.2 Myr. \label{fig1}}
\end{figure}


\begin{figure}
\epsscale{0.8}
\plotone{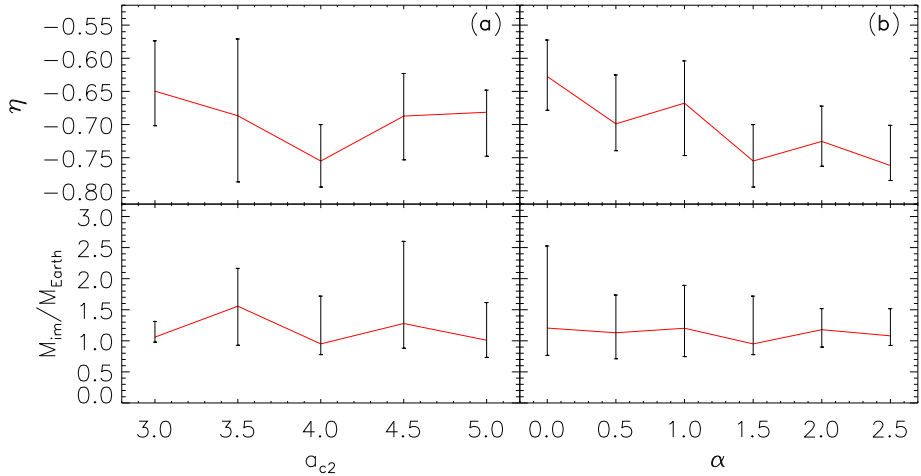}
\caption{Variation of the final $\eta=(a_{im,\,f} - a_{im,\,\,i})/a_{im,\,\,i}$ and the mass of the innermost embryos due to changes of the location of $a_{c2}$ (Panel a) or the surface density profile $\Sigma \propto {a^{ - \alpha }}$ (Panel b). Plotted are the median values at 1 Myr (20 runs for each point), the error bars are the upper and lower quartiles. \label{fig1}}
\end{figure}


\clearpage


\begin{thebibliography}{}
\bibitem[Artymowicz(1994)]{artymowicz94} Artymowicz, P., \& Lubow, S. H. 1994, \apj, 421, 651
\bibitem[Bromley \& Kenyon(2011)]{bromley11} Bromley, B. C., \& Kenyon, S. T. 2011, \apj, 735, 29
\bibitem[Chambers(1999)]{chambers99} Chambers, J. E. 1999, MNRAS, 304, 793
\bibitem[Cieza et al.(2009)]{cieza09} Cieza, L. A., Padgett, D. L., Allen, L. E., et al. 2009, \apjl, 696, L84
\bibitem[Doyle et al.(2011)]{doyle11} Doyle, L. R., Carter, J. A., Fabrycky, D. C., et al. 2011, Science, 333, 1602
\bibitem[Fragner \& Nelson (2010)]{fragner10} Fragner, M. M., \& Nelson, R. P. 2010, A\&A, 511, A77
\bibitem[Goldreich \& Tremaine(1979)]{Gold79}Goldreich, P., \& Tremaine, S., 1979, \apj, 233, 857
\bibitem[Hayashi(1981)]{hayashi81} Hayashi, C. 1981, Prog. Theor. Phys. Suppl., 70, 35
\bibitem[Holman \& Wiegert(1999)]{holman99} Holman, M. J., \& Wiegert, P. A. 1999, \aj, 117, 621
\bibitem[Ida et al.(2000)]{ida00} Ida, S., Bryden, G., Lin, D. N. C. \& Tanaka, H. 2000, \apj, 534, 428
\bibitem[Kasting et al.(1993)]{kasting93} Kasting, J. F., Whimire, D. P., \& Reynolds, R. T. 1993, Icarus, 101, 108
\bibitem[Kley \& Nelson(2012)]{kley12} Kley, W., \& Nelson, R. P. 2012, Ann. Rev. Astron. Astrophys., 50, 211
\bibitem[Kokubo \& Ida(2002)]{kokubo02} Kokubo, E., \& Ida, S. 2002, \apj, 666, 680
\bibitem[Kortenkamp et al.(2001)]{kortenkamp01} Kortenkamp, S. J., Wetherill, G. W., \& Inaba, S. 2001, Science, 293, 1127
\bibitem[Marzari et al.(2008)]{marzari08} Marzari, F., Th\'{e}bault, P., \& Scholl, H. 2008, \apj, 681, 1599
\bibitem[Marzari et al.(2012)]{marzari12} Marzari, F., Picogna, G., Desidera, S., \& Vanzani, V. 2012, Lunar and Planetary
Institute Science Conference Abstracts, Vol., 43, 1093
\bibitem[Meschiari(2012a)]{meschiari12a} Meschiari, S. 2012a, \apj, 752, 71
\bibitem[Meschiari(2012b)]{meschiari12b} Meschiari, S. 2012b, \apjl, in press (arXiv: 1210.7757)
\bibitem[Mischna et al.(2000)]{mischna00} Mischna, M. A., Kasting, J. F., Pavlov, A., \& Freedman, R. 2000, Icarus, 145, 546
\bibitem[Moriwaki \& Nakagawa(2004)]{moriwaki04} Moriwaki, K., \& Nakagawa, Y. 2004, \apj, 609, 1065
\bibitem[Ormel(2012)]{ormel12} Ormel, C. W., Ida, S., \& Tanaka, H. 2012, \apj, 758, 80
\bibitem[Orosz et al.(2012a)]{orosz12a} Orosz, J. A., Welsh, W. F., Carter, J. A., et al. 2012, \apj, 758, 87
\bibitem[Orosz et al.(2012b)]{orosz12b} Orosz, J. A., Welsh, W. F., Carter, J. A., et al. 2012, Science, 337, 1511
\bibitem[Paardekooper et al.(2012)]{paardekooper12} Paardekooper, S-J., Leinhardt, Z. M., Th\'{e}bault, P., \& Baruteau, C.  2012, \apjl, 754, L16
\bibitem[Payne et al.(2009)]{payne09} Payne, M. J., Wyatt, M. C., \& Th\'{e}bault, P. 2009, MNRAS, 400, 1936
\bibitem[Pierens \& Nelson(2007)]{pierens07} Pierens, A., \& Nelson, R. P. 2007, A\&A, 472, 993
\bibitem[Pierens \& Nelson(2008)]{pierens08} Pierens, A., \& Nelson, R. P. 2008, A\&A, 478, 939
\bibitem[Quarles et al.(2012)]{quarles12} Quarles, B., Musielak, Z. E., \& Cuntz, M. 2012, \apj, 750, 14
\bibitem[Quintana \& Lissauer(2006)]{quintana06} Quintana, E. V., \& Lissauer, J. J.  2006, Icarus, 185, 1
\bibitem[Scholl et al.(2007)]{scholl07} Scholl, H., Marzari, F., \& Th\'{e}bault, P. 2007, MNRAS, 380, 1119
\bibitem[Th\'{e}bault(2006)]{thebault06} Th\'{e}bault, P., Marzari, F., \& Scholl, H. 2006, Icarus, 183, 193
\bibitem[Ward(1997)]{Ward97}Ward, W. R. 1997, Icarus, 126, 261
\bibitem[Welsh et al.(2012)]{welsh12} Welsh, W. F., Orosz, J. A., Carter, J. A., et al. 2012, Nature, 481, 475
\bibitem[Walsh et al.(2012)]{walsh12} Walsh, K. J., Morbidelli, A., Raymond, S. N., et al. 2012, Nature, 475, 206
\end{thebibliography}
\end{document}